\documentclass[12pt,english]{article}
\usepackage[T1]{fontenc}
\usepackage[latin1]{inputenc}
\usepackage{a4}
\usepackage{babel}
\usepackage{graphics}
\usepackage{setspace}
\makeatletter

\providecommand{\LyX}{L\kern-.1667em\lower.25em\hbox{Y}\kern-.125emX\@}
\newcommand{\noun}[1]{\textsc{#1}}

 \newcommand{\lyxaddress}[1]{
   \par {\raggedright #1 
   \vspace{1.4em}
   \noindent\par}
 }

\makeatother
\begin{document}

\title{\noindent A microscopic model for exchange bias}

\author{A. Misra, U. Nowak and K. D. Usadel}

\maketitle

\lyxaddress{Theoretische Tieftemperaturphysik, Gerhard-Mercator-Universität Duisburg,
47048 Duisburg, Germany.}

\begin{abstract}
The domain state model for exchange bias has been further investigated
by considering vector spins for the antiferromagnet instead of Ising
spins used in the earlier studies. The qualitative results are similar
to those with infinite anisotropy for the antiferromagnet. However,
under certain conditions softer spins can lead to an even stronger
bias field. The study shows a nontrivial dependence of the exchange
bias on the antiferromagnetic anisotropy.
\end{abstract}

The shift of the hysteresis loop along the field axis in a system
containing a ferromagnet(FM)-antiferromagnet(AFM) interface is known
as exchange bias (EB). The discovery of this effect by Meiklejohn
and Bean in Co-CoO system \cite{mb56} dates back to 1956. Subsequently
the effect was observed in a variety of different materials (see \cite{ns99}
for a review). There have been several approaches towards a theoretical
understanding of the phenomenon and numerous different mechanisms
are believed to be responsible for the shift. Malozemoff \cite{m88}
explained the shift as a result of interface roughness. Koon considered
a spin-flop coupling between the FM and the compensated AFM interface
as responsible for the effect \cite{k97}. However, Schulthess and
Butler argued that the spin-flop coupling alone cannot account for
the effect \cite{sb98} but an interface coupling through uncompensated
defects together with the perpendicular coupling could explain the
mechanism \cite{sb99}.

Recently a model was proposed \cite{mgkbgnu00} where the cause of
the shift was attributed to the extra exchange field provided by the
remanent magnetization of the domains in a diluted AFM. The model
explained successfully several aspects related to the effect. The
essential idea behind the model is that when the diluted AFM is cooled
below the N\( \acute{\textrm{e}} \)el temperature in presence of
a field, it ends up in a domain state as opposed to long range order.
These domains are metastable since the domain walls require large
thermal fluctuations for any subsequent movement. The AFM is diluted
by replacing its atoms by nonmagnetic impurities (e.g. \( \textrm{Co}_{1-x}\textrm{Mg}_{x}\textrm{O} \)).
The domains are usually pinned at the impurity sites because at those
sites it does not cost any energy for the creation of a domain wall.
These domains carry a remanent magnetization \cite{nu91, k93, b98}
which provides the biasing field to the FM causing the shift of the
hysteresis loop. Several issues related to EB, such as temperature
dependence, training effect \cite{nu_u}, role of AFM thickness \cite{nmu01}
and, of course, the dependence of dilution on EB have been successfully
explained within the framework of this model. In the earlier work
\cite{mgkbgnu00, nu_u, nmu01}, owing to the strong uniaxial anisotropy
of CoO, the AFM was described by an Ising model. In the present work,
we relax this restriction on the AFM. In order to investigate a broad
class of systems for the AFM a Heisenberg model with variable uniaxial
anisotropy is considered. 

The system we will study consists of one monolayer of FM and \( t \)
monolayers of AFM. A certain fraction \( p \) of the AFM sites are
left without a spin. The FM is exchange coupled to the topmost layer
of the AFM. The Hamiltonian is thus given by \noun{\begin{eqnarray*}
\mathcal{H} & = & -J_{\mathrm{FM}}\sum _{\left\langle i,j\right\rangle }\overrightarrow{S_{i}}\cdot \overrightarrow{S_{j}}-\sum _{i}\left( d_{z}S^{2}_{iz}+d_{x}S_{ix}^{2}+\overrightarrow{S_{i}}\cdot \overrightarrow{B}\right) \\
 &  & -J_{\mathrm{AFM}}\sum _{\left\langle i,j\right\rangle }\epsilon _{i}\epsilon _{j}\overrightarrow{\sigma _{i}}\cdot \overrightarrow{\sigma _{j}}-\sum _{i}\epsilon _{i}\left( k_{z}\sigma ^{2}_{iz}+\overrightarrow{\sigma _{i}}\cdot \overrightarrow{B}\right) \\
 &  & -J_{\mathrm{INT}}\sum _{\left\langle i,j\right\rangle }\epsilon _{j}\overrightarrow{S_{i}}\cdot \overrightarrow{\sigma _{j}}
\end{eqnarray*}
} where \( \overrightarrow{S_{i}} \) and \( \overrightarrow{\sigma _{i}} \)
denote spins at the \( i \)th site corresponding to FM and AFM respectively.
The first line of the Hamiltonian describes the contribution to the
energy coming from the FM with the \( z \)-axis as its easy axis
\( \left( d_{z}>0\right)  \) and the \( x \)-axis as its hard axis
\( \left( d_{x}<0\right)  \). The in-plane anisotropy keeps the FM
preferentially in the \( y-z \) plane. The second line is the contribution
from the AFM with quenched disorder \( \left( \epsilon _{i}=0,\: 1\right)  \)
also having its easy axis along \( z \) \( \left( k_{z}>0\right)  \).
The last term describes the interaction of the FM with the interface
AFM monolayer. The whole system is placed in an external magnetic
field \( \overrightarrow{B}=B\widehat{z} \). In our simulations we
set \( J_{\mathrm{FM}}=-2J_{\mathrm{AFM}}=2J_{\mathrm{INT}} \). 

We use Monte Carlo methods with a heat bath algorithm. Each spin is
at every Monte Carlo step subject to a trial step consisting of a
small deviation from the original direction followed by a total flip.
This two-fold trial step can take care of a broad range of anisotropies
starting from very soft spins up to the Ising limit \cite{n01}. The
lateral dimension of the system was chosen in such a way that multiple
AFM domains can be observed which is the case for a lattice size \( 128\times 128\times (t+1) \).
Starting from a temperature above the N\( \acute{\textrm{e}} \)el
temperature \( \left( T_{N}\right)  \) of the diluted AFM but below
the Curie temperature \( \left( T_{c}\right)  \) of the FM, the system
is cooled below \( T_{N} \) in presence of an external magnetic field
\( \overrightarrow{B}=0.25J_{\mathrm{FM}}\widehat{z} \). Then the
hysteresis curve of the system is calculated along the \( \widehat{z} \)
direction. The EB field is determined as \( B_{\mathrm{EB}}=\left( B^{+}+B^{-}\right) /2, \)
where \( B^{+} \) and \( B^{-} \) are the values of the field where
the \( z \)-component of the magnetization of the FM vanishes when
the field is decreasing and increasing, respectively. 

From the simulations the following qualitative results emerge. In
absence of any anisotropy in the FM we observe a perpendicular coupling
between FM and AFM, only at very low dilution of the latter. The magnetization
reversal in the FM is by coherent rotation. The picture changes with
increasing uniaxial anisotropy in the FM and upon further dilution
of the AFM. The magnetization reversal in the FM is now by domain
wall motion and the perpendicular coupling becomes less significant.
This is because uniaxial anisotropies in both the FM and the AFM having
the same axis no longer lead to an energy minimum at perpendicular
coupling across the interface. Moreover AFM spins with missing AFM
neighbours can lower their energy by rotating parallel to the its
FM neighbour. Therefore, in the framework of our calculations a spin-flop
coupling is not an important mechanism for the shift of the hysteresis
loop.

Coloured pictures of typical staggered domain configurations of the
bulk AFM with different anisotropies can be found in the following
link: http://www.thp.uni-duisburg.de/\textasciitilde{}arko/picture.html.
The snapshots are taken when the net magnetization of the FM is nearly
zero during the decreasing field part of the hysteresis loop. As expected
the domain walls are wider for smaller values of \( k_{z} \), whereas
Ising like domain structures are obtained for stronger anisotropies.
Typical hysteresis loops for the FM and the interface monolayer of
the AFM are shown in Fig. 1. The upward shift of the hysteresis loop
for the interface layer of the AFM proves the existence of remanent
magnetization in the AFM domains. The magnitude of the EB field strongly
depends on the amount of this upward shift.

We have calculated the EB field for a wide range of values of \( k_{z} \),
starting from very soft spins to rigid, Ising-like spins. Fig. 2 shows
result for different thicknesses of the AFM for two values of the
dilution, \( p=0.4 \) and \( 0.6 \). A logarithmic scale is used
along the \( k_{z} \) axis only to increase the clarity of the figure.
Fig. 2(a) shows that for \( p=0.4 \) a thick AFM produces a peak
in the EB field at an intermediate value of \( k_{z} \) while at
lower thicknesses of the AFM the EB field increases with the anisotropy
and saturates in the Ising limit. Qualitatively this can be understood
as follows. The AFM domains are required to carry surplus magnetization
at the interface along the \( \widehat{z} \) direction which must
be stable during the hysteresis loop in order to produce any EB. At
low values of \( k_{z} \) the minimum energy configuration of the
AFM spins is predominantly dictated by the FM and the external field.
In this limit the AFM will follow the FM, thereby producing no bias
field. By increasing the anisotropy the AFM spins are forced to lie
more along the \( z \)-axis and hence the bias increases. However,
there exists a counter effect. Imagine a domain is formed upon field
cooling which preferentially goes through defects thereby minimizing
the energy. If now the anisotropy \( k_{z} \) is increased the domain
wall width will decrease and the energy to create such a wall will
increase. Thus the system will respond by flattening the domain boundaries.
This reduces the probability for the domains to carry any surplus
magnetization and hence the bias decreases for larger values of \( k_{z} \).
The compromise between these two opposite effects is achieved at some
intermediate value of \( k_{z} \) where the bias shows a peak. However,
the peak disappears at lower values of \( t \) (fig. 2(a)) or at
a higher dilution (fig. 2(b)). This happens when we are close to the
percolation threshold where the domain walls pass nearly exclusively
through the defects costing very little or no energy. Since no smoothening
of the domain walls is required now to minimize the energy of domain
formation, the second mechanism discussed above does not come into
play. Hence exchange bias increases with \( k_{z} \) till it saturates
in the Ising limit. Since the percolation threshold in three dimensions
is higher than that in two dimensions, increasing the number of AFM
monolayers at \( p=0.4 \) we go below the threshold and the peak
in the bias field reappears.

In conclusion, we find that the domain state model for EB proposed
originally for the Ising AFM is not restricted to this limit. Rather
under certain combination of thickness and dilution of the AFM, the
softness of the AFM spins can lead to an even stronger bias field.
There are several mechanisms responsible for the bias field, such
as dilution, thickness and anisotropy of the AFM. Although a qualitative
understanding regarding the dependence of EB on these parameters has
been achieved, a quantitative study of the domain structure both at
the interface and in the bulk of the AFM would provide a deeper understanding
to the problem. 

This work has been supported by the Deutsche Forschungsgemeinschaft
through SFB 491 and Graduiertenkolleg 277.

\newpage
{\centering \textbf{\Large Figure Captions}\Large \par}

Fig. 1. Typical hysteresis loop along \( \widehat{z} \) of (a) the
FM and (b) the interface monolayer of AFM, for \( p=0.6 \) and \( t=2 \).
The net magnetization is shown in units of the saturation magnetization
and the field in units of \( J_{\mathrm{FM}} \).

Fig. 2. Dependence of the exchange bias field, shown in units of \( J_{\mathrm{FM}} \),
on the AFM anisotropy for (a) \( p=0.4 \) and (b) \( p=0.6 \) at
different thicknesses.
\newpage

\vspace{0.5001cm}
{\centering \includegraphics{fig1.eps} \par}
\vspace{0.5001cm}

{\centering Figure 1: Misra et al.\par}

\newpage

\vspace{0.5001cm}
{\centering \includegraphics{fig2.eps} \par}
\vspace{0.5001cm}

{\centering Figure 2: Misra et al.\par}

\end{document}